%% file: main_v8.tex
\documentclass[10pt,comsoc]{IEEEtran}
\IEEEoverridecommandlockouts
\usepackage{cite}
\usepackage{amsmath,amssymb,amsfonts,helvet}
\usepackage[cmintegrals]{newtxmath}
\usepackage{bbm}
\usepackage{algorithmic}
\usepackage{graphicx}
\usepackage{textcomp}
\usepackage{soul}
\def\BibTeX{{\rm B\kern-.05em{\sc i\kern-.025em b}\kern-.08em
    T\kern-.1667em\lower.7ex\hbox{E}\kern-.125emX}}

\usepackage[
  pass,
]{geometry}
\usepackage{graphicx,epsfig}
\usepackage{subcaption}
\usepackage{caption,setspace}
\usepackage{cite}
\usepackage{url}
\usepackage[abs]{overpic}
\usepackage{footmisc}
\usepackage[alwaysadjust]{paralist}
\usepackage{hyperref}
\usepackage{multirow}
\usepackage{color, colortbl}
\usepackage{soul}
\usepackage{listings}
\usepackage{color}
\usepackage{makecell}
\usepackage{datetime}
\usepackage{pifont}
\usepackage{acro}
\input{acronyms.tex}
\usepackage{xcolor}

\newcommand{\red}[1]{\textcolor{red}{#1}}

\begin{document}

	\thispagestyle{empty}
	
	\twocolumn[
	\begin{@twocolumnfalse}
		{
			\vspace{1cm} 
			\noindent
			\textbf{Disclaimer:} This work has been accepted for publication in the IEEE Communications Magazine. © 2024 IEEE. Personal use of this material is permitted. Permission from IEEE must be
			obtained for all other uses, in any current or future media, including reprinting/republishing this material for advertising or promotional purposes, creating new collective works, for resale or redistribution to servers or lists, or reuse of any copyrighted	component of this work in other works.
			\vspace{1cm} 
		}
	\end{@twocolumnfalse}
	]

\title{Site-Specific Beam Alignment in 6G \\ via Deep Learning}

\author{Yuqiang~Heng,
        Yu~Zhang,
        Ahmed~Alkhateeb
        and Jeffrey~G.~Andrews
\thanks{Yuqiang Heng was and Jeffrey G. Andrews is with The University of Texas at Austin, Austin, TX, USA. Y. Heng is now with Samsung.  Yu Zhang and Ahmed Alkhateeb are with Arizona State University, Tempe, AZ, USA.}
\thanks{Corresponding author: J. G. Andrews (e-mail: jandrews@ece.utexas.edu).   Last modified: \today. }
}

\maketitle
	\setcounter{page}{1}

\begin{abstract}
\Acf*{BA} in modern millimeter wave standards such as 5G NR and WiGig (802.11ay) is based on exhaustive and/or hierarchical beam searches over pre-defined codebooks of wide and narrow beams.  This approach is slow and bandwidth/power-intensive, and is a considerable hindrance to the wide deployment of millimeter wave bands.  A new approach is needed as we move towards 6G.  \Acs*{BA} is a promising use case for \acf*{DL} in the 6G air interface, offering the possibility of automated custom tuning of the BA procedure for each cell based on its unique propagation environment and \acf*{UE} location patterns.   We overview and advocate for such an approach in this paper,  which we term \acf*{SSBA}.  \acs*{SSBA} largely eliminates wasteful searches and allows \acsp*{UE} to be found much more quickly and reliably, without many of the drawbacks of other \acl*{ML}-aided approaches.   We first overview and demonstrate new results on \acs*{SSBA}, then identify the key open challenges facing \acs*{SSBA}.
\end{abstract}

\section{Introduction}
Next generation cellular networks will need to deliver extremely high data rates for emerging applications, which will necessitate much more effective utilization of the vast amount of spectrum above 28 GHz.  The high isotropic pathloss at these frequencies require highly directional \ac{BF}, where \acp{BS} and \acp{UE} -- both equipped with dense antenna arrays -- focus energy in particular directions. Finding and maintaining near-optimal \ac{BF} directions -- a process known as \ac{BA} -- is the critical bottleneck to unleashing this spectrum.   Done correctly, high directionality also provides a path to much improved power efficiency, which will be required for important emerging use cases such as Augmented Reality (AR) glasses that have both high bandwidth demands and small power budgets.

The \ac{BA} framework in 5G relies on extensive beam sweeping, measurement and reporting. In the downlink, the \ac{BS} periodically sweeps one or more generic codebooks of pre-defined beams by transmitting beamformed \acp{RS} while the \ac{UE} sweeps its \ac{Rx} codebook\footnote{Although this article is focused on cellular networks, in particular 5G and 6G, nearly everything we discuss is directly applicable to other millimeter wave systems such as the various WiGig standards operating in the 60 GHz band, which also rely on beam sweeping and suffer from slow \ac{BA}.}.
The best beam or multiple beam pairs with the highest received power are selected and reported to the \ac{BS}.
This simple procedure handles the \ac{BA} procedure both for \ac{IA} of previously undetected UEs and for tracking already connected UEs.  To guarantee that most unconnected \acp{UE} can be found during \ac{IA}, generic codebooks with quantized \ac{BF} angles that cover the entire angular space are usually adopted.  While on some level ``foolproof'' since new UEs can be found regardless of their location in the angular space, this approach to \ac{BA} is obviously inefficient.   It effectively starts from scratch during each search cycle, learning nothing from previous searches.  It is agnostic to the propagation environment or the historic probability of finding a UE in a particular direction.

For these reasons, a data-driven and learning-based approach should be highly beneficial for \ac{BA} for 6G. In particular, we will argue that \ac{DL} is well-suited to tackle these challenges with its powerful function approximation capabilities, relatively low complexity, and its well understood training and convergence properties.  By leveraging characteristics of the propagation environment and patterns of \ac{UE} distribution and mobility, a site-specific DL approach can eliminate unproductive searches while quickly predicting beams that point accurately towards \acp{UE}.

While a few recent works have provided useful surveys of general applications of \ac{DL} in beam management \cite{khan2023DL_BM_survey,ma2022DL_BM_survey}, we present a more focused perspective by limiting our scope to the spatial and site-specific aspects.    We first identify (Sect. \ref{sect:DL-BA_criteria}) the key requirements of \ac{BA} and their implication to \ac{DL}-based approaches. We then overview the concept and key aspects and advantages of state-of-the-art of site-specific \ac{DL} techniques for \ac{BA} in Sect. \ref{sect:site-specific}.   In Sect. \ref{sect:EE_approaches} we overview 3 specific ways to do \acf{SSBA} that have been developed independently by the authors, and we present a unified and novel comparison of their performance using ray tracing in a Boston neighborhood.  We include other baselines and theoretical upper bounds and observe the consistently immense potential of \ac{SSBA} for improving beamforming gain with a drastically reduced number of beam searches. A number of important open problems remain, and these challenges are identified and several promising directions for future research are proposed to conclude the article.

\section{Key Criteria for DL-aided BA}\label{sect:DL-BA_criteria}
An intelligent \ac{BA} method should be able to accurately and quickly identify high \ac{SNR} beams for each UE in the cell without exhaustively searching all candidates. However, high \ac{BF} gain and low search latency are not the only requirements of an ideal \ac{BA} method. 
After all, the exhaustive and the hierarchical search with uniform codebooks have been adopted in 5G largely because they can be deployed in any cell site without any custom adaptation, and they allow new \acp{UE} to be discovered without special side information (e.g. GPS coordinates) or excessive amounts of feedback.

Therefore, we first identify the key requirements for a 6G \ac{BA} method.   These four requirements should apply universally, and in many cases they rule out proposed \ac{DL}-aided approaches, as we now discuss.

\textbf{R1: Accurate and Fast over Entire UE population.}  A trade-off between speed and \ac{SNR} is unavoidable in most \ac{BA} methods: better beams may be found by increasing the resolution of the codebook or more frequently sweeping the codebook, at the direct expense of latency and overhead.   However, some BA methods may present much better such SNR-speed tradeoffs than others, especially when BA for all UEs is considered collectively.  Notably, many proposed \Ac{DL}-based approaches leverage environment-specific and even \ac{UE}-specific features, but the gain diminishes about linearly with the number of \acp{UE} due to the requirement of a per-\ac{UE} search. This is because the beam refinement search in 5G NR has to be conducted through \ac{CSI-RS} on a per-\ac{UE} basis even though the \acp{SSB} beams are broadcasted cell-wide \cite{heng2021BM_magazine}. Therefore, the \emph{total cell-wide latency}  -- the number of searches to achieve \ac{BA} for \underline{all} \acp{UE} -- is the correct latency metric.

\textbf{R2: Versatile.}  A \ac{BA} method should work well in a large variety of deployments, including urban, suburban, outdoor and indoor environments with many different types of \acp{UE}. Each deployment has unique challenges: an outdoor vehicular \ac{UE} has high velocity but predictable mobility patterns while an indoor \ac{XR} user may experience fast unpredictable rotations and more frequent \ac{NLOS} conditions. Rather than designing a different \ac{BA} approach for each scenario, e.g. using application-specific sensor data, a desirable \ac{BA} method should handle a wide range of scenarios in a single framework.  

\textbf{R3: Scalable to Higher Carrier Frequencies.} While the array size and thus channel dimensionality will grow considerably as 6G moves up in the spectrum, the \ac{BA} complexity and latency should only increase moderately. This is achievable in theory since channel sparsity is preserved at higher carrier frequencies.  \ac{DL}-based methods can learn the underlying channel structure and intelligently predict the optimal beams without significantly increasing the \ac{BA} latency. Grid-free approaches that directly compute \ac{BF} weights will become more attractive since codebooks with thousands of narrow beams will prove cumbersome.

\textbf{R4: Self-training and Auto-updating.} For real-world adoption of \ac{DL} in \ac{BA}, the \ac{BS} and \acp{UE} could be deployed as usual with a default codebook and/or \ac{BA} procedure. Learning should ideally utilize ongoing over-the-air measurements possibly along with mostly automated site-specific simulation, so that devices can seamlessly transition to perform \ac{BA} with the \ac{DL}-aided methods once they have been sufficiently trained. Furthermore, the \ac{DL} models should also be able to continuously (or at least periodically) improve and adapt. This is a major open challenge not well-addressed to date, as most existing approaches assume extensive offline training prior to deployment and when or how to re-train is also not well understood.

\begin{figure*}[h!]
	\centering
	\includegraphics[width=.7\textwidth]{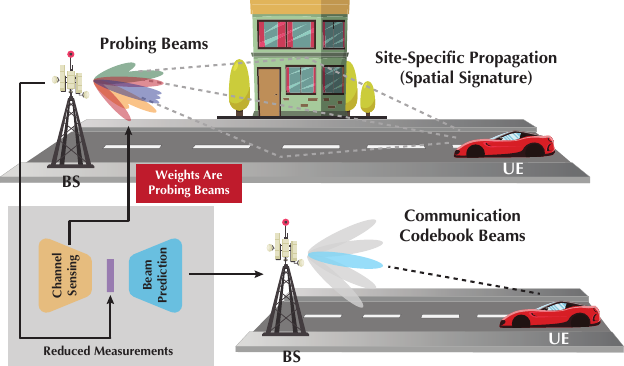}
    \caption{An illustration of SSBA that consists of channel probing and beam selection. The figure presents the deployment phase where the system can directly select the desired beam based on channel sensing power measurements.}
	\label{fig:sys}
\end{figure*}

\section{Beam Alignment with Site-Specific Learning}\label{sect:site-specific}
The \ac{BA} procedure has two main components: \textbf{channel sensing} and \textbf{beam selection}. The objective of channel sensing is to collect spatial information about the channel between the transmitter and receiver. This is done using a \textit{channel probing codebook} which could be as simple as a fixed set of relatively wide beams (as in 5G NR \ac{IA}) or a more sophisticated compressive sensing measurement codebook. The measurements from channel sensing are then leveraged to design the data transmission beam either by selecting it from a pre-defined codebook or by computing an arbitrary beam.  Interestingly, these two stages strongly depend on the attributes of the specific site and environment, such as the outdoor buildings' geometry or indoor floor layout, antenna panel orientation, and user locations.  This motivates what we call \textit{site-specific learning for \ac{BA}}, whereby both the sensing and the beam selection tasks attempt to exploit aspects of their specific cell site.  In this section, we outline the key ideas behind site-specific sensing and beam prediction and highlight both the potential gains and key design considerations.

\textbf{Traditional Approaches Are Not Site-Specific.}  Traditional \ac{BA} techniques generally follow one of two main approaches: \emph{beam training} or \emph{channel estimation-based}.  In beam training, the transmitter and receiver sweep over the beams in the probing codebook in an attempt to select the most promising pair of beams. This beam pair could be directly used for data transmission in the case of exhaustive search, or further refined, for example via hierarchical codebooks with different beamwidths. Beam training is the \ac{BA} approach adopted in IEEE 802.11ad/ay, as well as in at least the first three releases of 5G  \cite{heng2021BM_magazine}.  For channel estimation-based beam design, the \ac{MIMO} matrix channel is first estimated, typically using compressive sensing measurement codebooks but possibly via other channel estimation methods based on the reception of pilot (reference) signals \cite{heath2016overview}.   The estimated channel is then used to design the beamforming vectors, for example the maximum left (transmit) and right (receive) singular vectors of the channel matrix.  In neither case does the channel probing codebook and the beam selection process leverage site-specific attributes nor prior observations.

\textbf{Potential Gains with Site-Specific Optimization.} Since the two components of the \ac{BA} process -- channel sensing and beam selection -- rely heavily on characteristics such as the geometry of the buildings and scatterers around the \ac{BS} and \acp{UE}, as well as the \ac{UE} locations, it is intuitive that optimizing each \ac{BS}'s probing codebook and beam selection criteria based on these site-specific characteristics may significantly improve the \ac{BA} performance. For example, instead of scanning all directions, the probing beam codebook could focus on the most frequently useful directions and avoid directions with \ac{LOS} blockages or where UEs are rarely found.  Similarly, channel compressive sensing codebooks could be refined to focus  the sensing energy on the most important dimensional subspaces. The probing beams can be transmitted using \acp{SSB} in a 6G system, allowing new \acp{UE} to be discovered and complete \ac{IA}. Furthermore, beam selection can also leverage site-specific prior observations. Conceptually, it can bias towards more commonly selected beams and deprioritize beam pairs that have often provided low \ac{SNR} previously.  This approach can reduce the \ac{BA} overhead and enhance the beam selection accuracy as shown in Section \ref{sect:EE_approaches}.

\textbf{Role of Data and Deep Learning.} Realizing the potential of site-specific \ac{BA} optimization using classical signal processing techniques is non-trivial. First, mmWave systems rely on analog-only or hybrid analog/digital transceiver architectures which impose strict constraints on the beamforming weights. This makes the site-specific probing codebook optimization problem highly non-convex and difficult to solve. Second, unless the probing codebook has simple steering beams, the mapping function that maps the outcomes of the channel sensing to the best data transmission beam is hard to characterize analytically. This motivates leveraging data-driven \ac{DL}  to design site-specific \ac{BA} approaches. \Acp{DNN} possess superior expressive power and have been proven successful in solving many challenging, non-convex problems. In particular, with proper design of the \ac{ML} architecture, loss function, and learning strategy, we can learn both optimized site-specific channel probing codebooks and functions that map the measurements of these probing codebooks to data transmission beams.   

\textbf{Key Considerations for Site-Specific Learning.}   We emphasize two important considerations for \ac{SSBA}.  First, developing and evaluating site-specific beam prediction solutions requires using site-specific channel datasets, either from the real world or accurate ray-tracing simulations.  This is essential since the key idea here is that the probing codebook and the beam selection are based on the underlying channel structure of the specific deployment.  This per-site specialization is where the gains come from, and therefore using general statistical channel models that do not capture the dependency on the site geometry and user distribution will prove unsuccessful. We will argue that the ``price'' for this data acquisition and site-specific modeling is well-justified and not necessarily even very large, but nevertheless this is a key new challenge versus current approaches that are not site-specific.

Second, it is important to differentiate between the design and the training of the \ac{BA} machine learning model.  While the training should mainly utilize site-specific datasets and measurements, and thus be optimized and tuned for the specific deployment, the model itself should be universal and scalable so that it could work in a large number of sites across a variety of deployments.

\section{End-to-End Learning for Site-Specific Beam Alignment}\label{sect:EE_approaches}

As discussed in Section \ref{sect:site-specific}, the \ac{SSBA} has two sub-problems: (i) learning site-specific channel probing codebook and (ii) learning the mapping from the channel sensing measurements to beams. The two problems are coupled, where the ultimate objective is to achieve accurate prediction with smallest possible number of measurements.
Given this coupling, end-to-end learning that jointly optimizes the probing codebook and learns the mapping function based on a common loss function is our recommended approach for \ac{SSBA} solutions.
{In this section, we present two end-to-end learning frameworks for \ac{SSBA} and discuss their tradeoffs, and compare them in a common experimental setting.}

\begin{figure*}[hbt]
	\centering
	\includegraphics[width=0.85\textwidth]{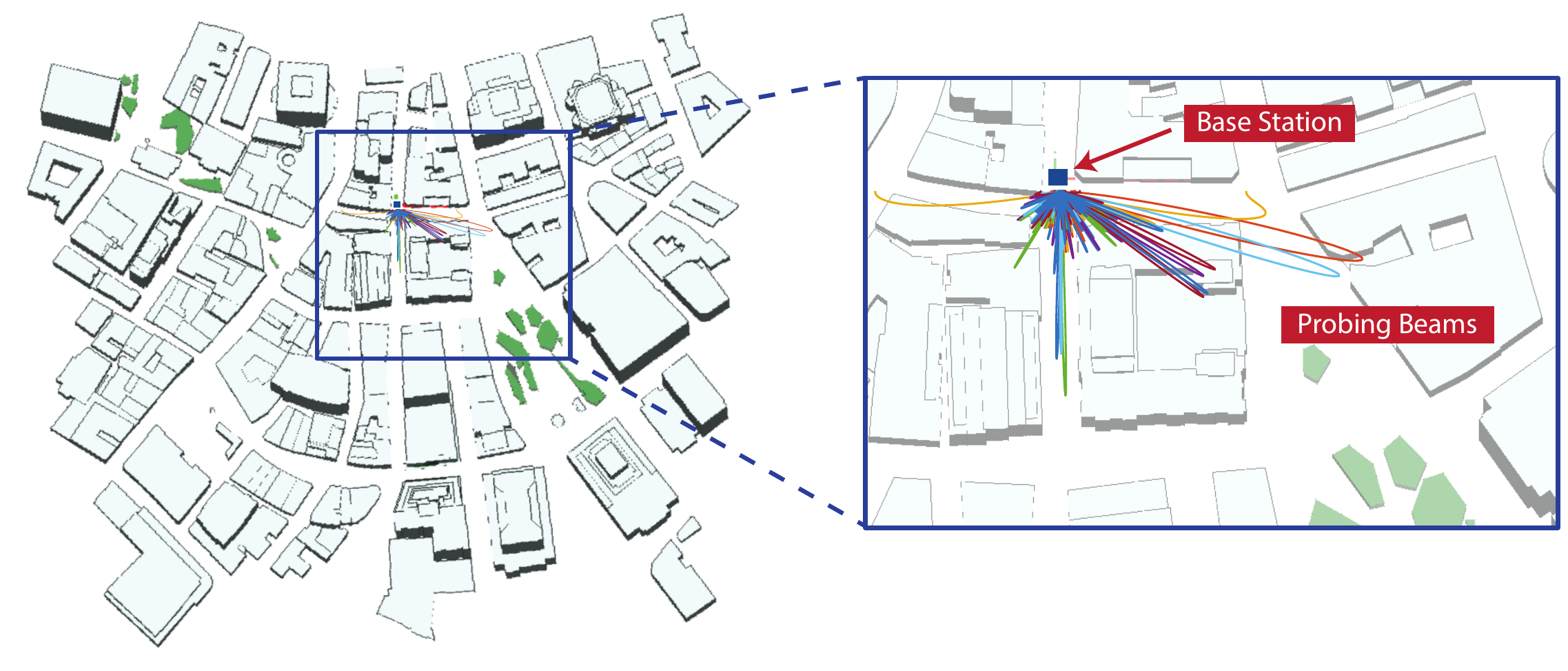}
	\caption{Illustration of DeepMIMO Boston5G scenario with the learned probing beam patterns overlaid for visualization purposes.}
	\label{fig:RT_env}
\end{figure*}

\begin{figure}[hbt]
	\centering
	\includegraphics[width=1\columnwidth]{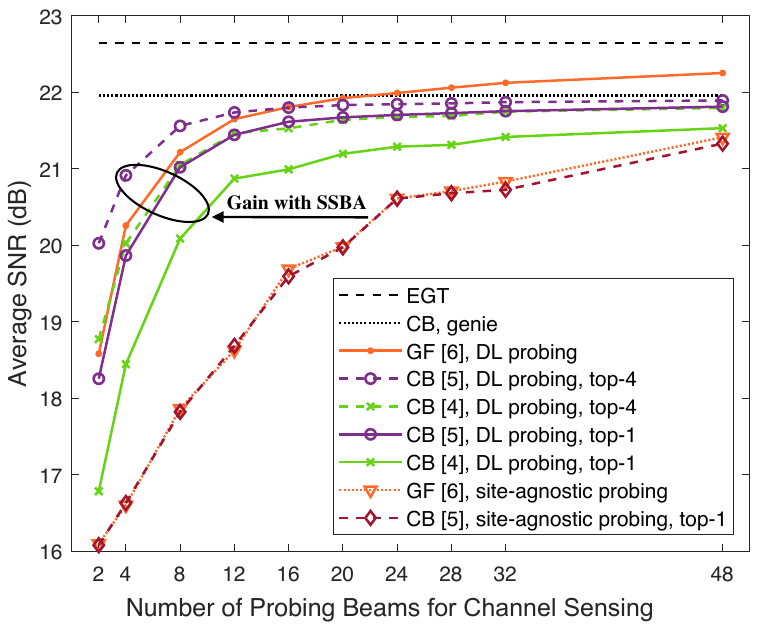}
	\caption{Comparison of the average SNR achieved by site-specific CB and GF approaches and site-agnostic baselines. }
	\label{fig:avg_SNR_vs_num_probe}
\end{figure}

\textbf{Codebook Based Beam Prediction.}  The first framework is based on selecting the best narrow beam from a discrete set, i.e. from a codebook.  The \ac{CB} approach is closely related to existing standards like 5G, and the option to perform over-the-air beam refinement makes its beam predictions more robust. Beam tracking is also more intuitive and convenient with \ac{CB} approaches, since adjacent beams in the codebook can be monitored.
We consider two \ac{CB}-based approaches, the first is adapted from the early work on site-specific BA in  \cite{li19hybrid} and the second is a more recent one, which is presented in full detail in \cite{Heng2022CB}.

\textbf{Grid-free Beam Prediction.} The \ac{GF} approach holds out the possibility of identifying a truly optimum beam, as well as eliminating any beam refinement search phases, which become quite costly in a network setting as they must be conducted on a per-\ac{UE} basis.  Thus, in principle, a well-functioning and well-trained \ac{GF} approach could potentially require far fewer total beam search measurements over multiple \acp{UE}.  However, the \ac{GF} approach is more risky, since a poorly calculated beam can be arbitrarily poor, and often the training and beam normalizations are more sensitive than for a predefined codebook.  The results may also be less interpretable since the action space is infinitely large. Our \ac{GF} approach is from \cite{Heng2023GF}.

\textbf{Unified Ray Tracing Experiment.}  We now consider a cellular simulation environment based on a section of downtown Boston, shown in Fig. \ref{fig:RT_env}. The \ac{BS} is located at an intersection. There are a total of 77,597 \acp{UE} -- of which 52\% are \ac{LOS} and 48\% are \ac{NLOS} -- located along the two horizontal and vertical streets closest to the \ac{BS}. The channels are generated through ray-tracing as discussed in \cite{alkhateeb2019deepmimo}, with simulation parameters summarized in Table \ref{table:sim_params}. In particular, we assume a $64 \times 1$ \ac{MISO} \ac{ULA} scenario, but these approaches do extend to more general \ac{MIMO} and \ac{UPA} models. The \ac{BS} performs analog \ac{BF} with unquantized phase shifters, for which \ac{EGT} with perfect per antenna phase alignment is a true upper bound. An over-sampled \ac{DFT} codebook with 256 beams is adopted for the codebook-based baselines. The \ac{CB} genie corresponds to the best beam in the codebook.

The first \ac{CB} approach modifies the \ac{NN} architecture originally proposed in \cite{li19hybrid}, perhaps the earliest work on end-to-end learning for site-specific BA. The probing codebook is designed via a complex-valued NN and parameterized with the phase, from which the real and imaginary parts of the probing beams are computed to enforce the unit-modulus constraint.     The mapping function consists of 2 hidden layers with 512 and 1024 neurons, each with ReLu activation followed by batch normalization, and an output layer with sigmoid activation. The \ac{NN} is trained to minimize the average binary cross-entropy between the model output and the one-hot encoded labels.

The second \ac{CB} approach is from \cite{Heng2022CB}. The probing codebook is implemented as a complex-valued linear layer, which is normalized per element to ensure the unit-modulus constraint. The mapping function consists of 2 hidden layers with 520 neurons each and ReLu activation, and a linear output layer with softmax activation. The \ac{NN} is trained to minimize the cross-entropy loss. Both \ac{CB} models output the predicted posterior probability of the optimal beam index.

The \ac{GF} approach adopts the \ac{NN} architecture in \cite{Heng2023GF}. The probing codebook is implemented as a complex-valued linear layer with per-element normalization. The mapping function consists of 2 hidden layers 520 neurons each and ReLu activation, and a final linear layer that outputs the real and imaginary parts of the predicted \ac{BF} vector. The predicted \ac{BF} vector is also normalized element-wise to ensure the unit-modulus constraint. The \ac{NN} is trained to maximize the average \ac{BF} gain normalized by the channel norm.

\begin{table}
\small
\centering
  \caption{Simulation Parameters}\label{table:sim_params}
    \begin{tabular}{| c | c |}
    \hline
    BS Antenna & $64\times1$ ULA\\ \hline
    BS Codebook Size & 256\\ \hline
    Antenna Element & Isotropic \\ \hline
    Carrier Frequency & 28 GHz\\ \hline
    Bandwidth & 50 MHz\\ \hline
    Transmit Power & 40 dBm\\ \hline
    Noise PSD & -161 dBm/Hz \\ \hline
    Probing Spreading Gain & 32 \\ \hline
    \end{tabular}
\end{table}

The average \ac{SNR} achieved by the site-specific (i.e. ``DL probing'') \ac{CB} and \ac{GF} approaches and the site-agnostic and idealized baselines are shown in Fig. \ref{fig:avg_SNR_vs_num_probe}.   It is important to remember that an exhaustive codebook search would entail 256 measurements and still perform below the \ac{CB} genie in terms of \ac{SNR} (due to noise and the corresponding measurement and feedback errors).  We summarize some of the key takeaways and insights now.

\begin{itemize}
    \item \textbf{Near optimal performance is achieved with a fraction of the measurements/latency}.  Both the \ac{GF} and \ac{CB} methods can achieve \ac{SNR} within 1 and 0.5 dB from that of the genie using just 8 and 16 probing measurements, respectively. 
    Compared to an exhaustive \ac{CB} search, the latency is reduced by $32\times$ and $16\times$, respectively. The power demand of the lightweight \ac{NN} should be easily accommodated by \acp{BS}, while \acp{UE} can save power by measuring fewer beams.
    \item \textbf{GF beats CB} in terms of total measurements.   By eschewing a refinement search phase -- which is very helpful to the \ac{CB} techniques -- the \ac{GF} approach holds out the ultimate promise for low latency searches.   This is even more true if the \ac{UE} performs beamforming as well.
     {\item \textbf{CB approaches could be more robust}. Since the top-k predicted beams could be refined \ac{OTA} (at additional overhead cost) to select the best beam, \ac{CB} approaches could generally be more robust to prediction errors and less sensitive to quick changes in the environment.}
    \item \textbf{Site-specific probing is indispensable}.  If the learned probing beams are replaced with evenly-spaced ``site agnositc'' narrow beams and the \acp{NN} are re-trained from scratch, there is a large performance loss, of over 3 dB or equivalently a factor of at least two in terms of required probing measurements.
\end{itemize}

The shape of the learned probing beams is illustrated in Fig. \ref{fig:RT_env}.  Note that a given probing beam can have several lobes of varying strength, but each probing beam has the same total transmit energy.   They are clearly adapted to the environment: they mainly focus their energy to cover the horizontal and vertical streets on which \acp{UE} are scattered.  This intuitively demonstrates the importance of \ac{SSBA}.

\section{Future Research Directions}
\label{sect:future_directions}

While these \ac{DL}-aided \ac{SSBA} solutions show great promise, there remain important open problems to solve.
In this section, we outline these important research problems and identify promising future directions. 

\subsection{Practical Training and Deployment Approaches}
Existing \ac{DL}-aided \ac{BA} methods typically employ an offline training phase prior to deployment and the models need to be retrained whenever the environment changes. Collecting the large amount of training data required to represent the site prohibits dynamic adaptation and network-wide deployment of the site-specific models. {Further, some \ac{DL}-aided \ac{BA} methods require explicit full channel knowledge in the training phase, which is hard to obtain in practice.}   Clearly, new deployment paradigms will be necessary for \ac{SSBA} to have a real-world impact.

One encouraging direction lies in the adoption of digital twins, which promise true-to-life simulations of the physical environment at large scales \cite{alkhateeb2023DT}. High fidelity 3D models of entire cities can be constructed from \ac{LiDAR} and satellite data, which can also be dynamically updated based on live-monitoring of the environment and \ac{UE} distribution.  Enabled by the vast computational power of the latest \acp{GPU} in the cloud, high quality channels of millions of \acp{UE} can be simultaneously generated with real-time ray-tracing. {Initial applications are already seen in 5G cell planning \cite{heavyRF} using Nvidia's Omniverse-based Aerial platform.}

\begin{figure*}[hbt]
   \centering
   \includegraphics[width=0.7\textwidth]{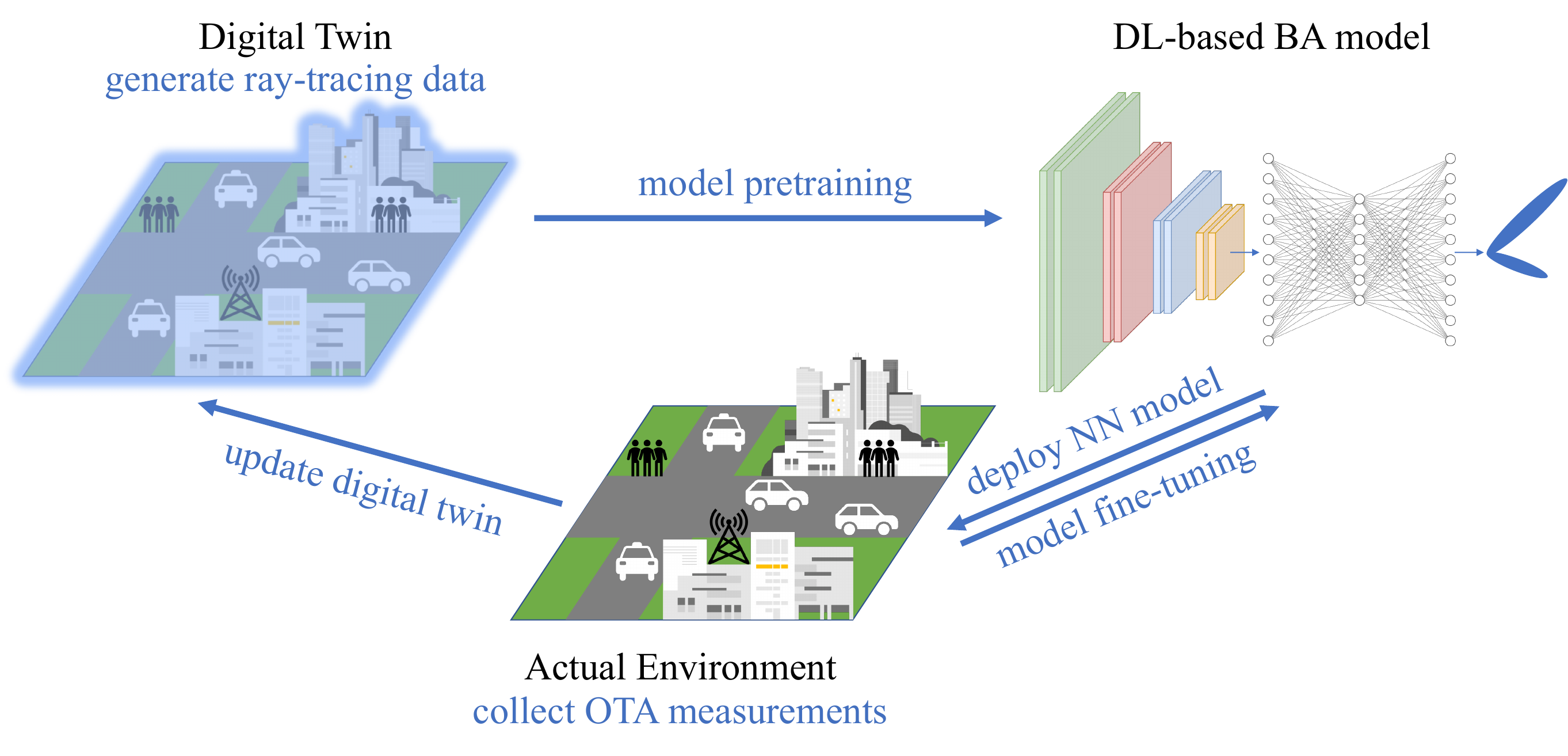}
    \caption{Illustration of a DL-based BA pipeline aided by digital twin. The digital twin provides ray-tracing data to pretrain the DL-based BA model. Measurements from the actual environment are used to fine-tune the BA model and update the digital twin.}
   \label{fig:digital_twin}
\end{figure*}

An envisioned use case for \ac{SSBA} is illustrated in Fig. \ref{fig:digital_twin}, where the digital twin keeps a dynamically updated model of a city-wide network. 
Copies of the \ac{SSBA} \ac{NN} models are constantly fine-tuned or re-trained using data continuously generated by the digital twin in the background, which can be rapidly deployed to all the \acp{BS} to replace the outdated models and calibrated through few-shot transfer learning with a few \ac{OTA} measurements.
Such digital twins are envisioned to be used for many other applications as well, including outdoor \ac{XR} experiences, driverless cars, and other types of communication optimization, and so the required data and platforms for \ac{SSBA} may be available nearly ``for free" in the 6G era.

\subsection{Advanced Deep Learning Approaches}

Existing work on \ac{DL}-aided BA focused mainly on developing efficient learning models that can perform well on small-scale datasets. Realizing the potential gains of \ac{SSBA}  in practical deployments, however, requires developing full \ac{ML} operation frameworks that actively select which data to keep, monitor distribution shifts, efficiently process large-scale datasets, and frequently update the learning of the \ac{SSBA} models \cite{Ovadia2019NeurIPS-Short}.  This is essential as the data-driven nature of \ac{SSBA} inherently requires these solutions to be able to detect and account for any channel distribution shifts.
Further, how to efficiently select and utilize past data/observations is also an important research direction. For instance, it might be promising to leverage continual learning concepts to develop systematic frameworks that can continuously acquire, update, accumulate, and exploit knowledge throughout the operation lifetime \cite{Friedemann2017ICML-Short}.

Another important direction is developing more advanced learning architectures that can scale to more complex systems and diverse deployment scenarios. Leveraging tools from active learning to iteratively generate probing beams based on previous measurements \cite{SohrabiYu2022ActiveSensingDL} is an interesting approach to further reduce latency compared to using a fixed set of probing beams. However, these approaches normally scale poorly with the number of \acp{UE} since the probing beams are \ac{UE}-specific. Future research should investigate the gain of site-specific and \ac{UE}-specific \ac{BA} and improve its scalability under practical network considerations. Multi-task learning is a potential approach to improve the learning and data efficiency while making models more robust and generalizable \cite{liu2019multitask}. Auxiliary tasks such as channel estimation and localization can be solved simultaneously in additional to beam alignment by designing a \ac{NN} with a common probing codebook for channel sensing and sub-modules to produce task-specific outputs.

\subsection{Coverage and robustness}
While \ac{DL}-aided approaches often demonstrate much faster \ac{BA}, they have not yet addressed concerns over robustness and reliability.
Metrics such as average \ac{SNR} and beam selection accuracy used by most \ac{DL} models do not paint the whole picture. Operators often care more about providing a minimal performance guarantee to cell-edge \acp{UE} and avoiding link failures than maximizing the \ac{SNR} for the top-percentile \acp{UE}. Future research needs to focus on the entire performance distribution and particularly on cell-edge \acp{UE}. Reliability metrics such as the coverage should also be incorporated into the training objective.

\ac{DL}-aided \ac{BA} methods also need to be robust against imperfections that arise in practical deployments. For example, the channels used during training may be different from the actual channel distribution experienced in practice, which could be caused by channel estimation error, channel model inaccuracy or mismatch between the simulated and actual environment. Noisy \ac{ESI} and \ac{RF} hardware imperfections will also deteriorate the \ac{BA} performance. Such imperfections need to be considered and modeled in the training process. Adversarial training can further improve the robustness of these \ac{DL} models and improve generalizability. Future research may also explore more transparent and interpretable \ac{DL} models, which would allow us to better assess their performance in practical scenarios.

\subsection{Uplink Alignment}
Existing research on \ac{DL}-aided \ac{BA} has largely focused on improving latency on the \ac{BS} side and has not addressed several unique challenges faced by \acp{UE}, namely power conservation and nontrivial mobility and rotations. Millimeter wave \acp{UE} in 5G are assumed to have relatively few beams and moderate mobility. In emerging use cases such as high-speed trains, unmanned aerial vehicle and \ac{XR}, \acp{UE} will experience more complex mobility patterns and faster velocity and rotations. Better tools are needed to model the 3D mobility, rotation pattern and effect of self-blockage of \acp{UE}.

Uplink \acp{RS} defined in 5G can be repurposed for uplink probing. The equivalent of \ac{SSBA} for the uplink should learn to capture additional information such as the rotation pattern, correlation between antenna panels and device self-blockage pattern that are unique to the \ac{UE}. {This could lead to joint site and \ac{UE}-specific \ac{BA} approaches.} 
As \ac{BF} is intricately tied to the characteristics of \acp{UE}, influenced by hardware and implementation variations, the ideal probing codebook and beam selection function may be different for each \ac{UE}. Ensuring the generalizability of \ac{SSBA} to a range of \acp{UE}, or its ability to evolve alongside shifts in \ac{UE} populations, is crucial.
In general, if the impressive reduction in the number of downlink measurements achieved by \ac{SSBA} can be translated to the uplink, the potential for power and latency reduction is huge.

\subsection{Network-wide Multi-cell Optimization}
Existing approaches have considered single-cell scenarios with uniformly distributed or clusters of \acp{UE}. Future research may explore optimization in a multi-cell network.
Whereas cell boundaries in 5G are determined by the strongest beams in the uniform codebooks, \ac{DL}-aided \ac{BA} methods may learn site-specific codebooks or even \ac{GF} \ac{BF}. Hence, network-wide optimization needs to coordinate among cells, each providing non-uniform coverage. Training in a multi-cell deployment can also take place in a centralized or distributed fashion, where federated learning is a promising tool \cite{elbir2021federated}. For instance, the centralized cloud may gather data across the network to learn a large model, which is distilled and adapted to each individual cell using the smaller-scale site-specific data. Finally, neighboring cells may share elements of their environment, especially in denser networks at higher carrier frequencies. Combined with the rising interest in multi-point connectivity, information sharing and coordination \ac{BA} among nearby \acp{BS} is another promising research direction. {In dense networks, the \ac{BA} decisions may also be coupled with other user association and hand-off decisions, which motivate developing multi-task \ac{DL} models to address these cases.}

\subsection{Standardization and Commercial Deployment}

Standardization of \ac{BA} in 5G is built around assumptions of sweeping, measurement and reporting of codebooks of beams. For instance, the \ac{BS} and the \ac{UE} establish a mutual correspondence between beams in the codebook, \acp{RS} and resource blocks through \ac{TCI} states. On the other hand, \ac{DL}-aided methods can use enormous codebooks represented by \acp{DNN}, making the existing codebook-based signaling cumbersome.
Furthermore, the beams used by the \ac{UE} are transparent to the \ac{BS} during feedback in 5G. For \ac{BA} methods that rely on in-band sensing, this limits the information available to the \ac{BS}, which will be a severe bottleneck when both \ac{BS} and \ac{UE} have large antenna arrays.
Future standards should dedicate resources and signals for learning, such as for testing the learned beams before deployment and for more robust and prioritized feedback of the \ac{ESI} or sensing measurements.
While the existing \ac{BA} framework is simple, we need to reconsider codebook-based assumptions and allow for more flexibility to accommodate powerful \ac{DL}-aided \ac{BA} methods.

To accelerate research on \ac{DL}-aided \ac{BA}, more complete and diverse public datasets should be developed for easier training and standardized performance benchmarks.  Competitions should be hosted with a variety of hidden testing data to encourage more competitive \ac{DL} models and faster design iterations. The industry can also accelerate \ac{SSBA} by providing practical deployment scenarios,  allowing use of commercial-grade simulators, and sharing real-world measurement data. Deciding when these \ac{DL}-aided \ac{BA} techniques are ready for actual deployment is another challenge. 

\section{Conclusion}

\ac{SSBA} is an exciting application of deep learning that can provide an increasingly rare opportunity for order-of-magnitude improvements in physical layer performance metrics.   It is a promising use case for \ac{DL} in 5G-Advanced and 6G, and could prove to be a crucial enabler for wider coverage and deployment of mmWave spectrum, if the challenges identified in this article can be satisfactorily addressed.

\vspace{5mm}

\noindent \textbf{Yuqiang Heng} is a senior research engineer at Samsung. He earned his PhD at UT Austin in 2022 and BS at Rice. \vspace{2mm}

\noindent \textbf{Yu Zhang} received his BS and MS from Beijing Jiaotong University. He is a PhD student at ASU. \vspace{2mm}

\noindent \textbf{Ahmed Alkhateeb} is an Assistant Professor at ASU. He holds a PhD from UT Austin and has received awards including the 2016 IEEE Signal Processing Society Young Author Best Paper Award. \vspace{2mm}

\noindent \textbf{Jeffrey Andrews} is the Truchard Family Endowed Chair in Engineering at UT Austin and the Director of 6G@UT. He received the 2019 IEEE Kiyo Tomiyasu Award and holds a PhD from Stanford.\vspace{2mm}

\end{document}

%% file: acronyms.tex
\DeclareAcronym{3GPP}{
  short=3GPP,
  long=3rd generation partnership project
}
\DeclareAcronym{ADC}{
  short=ADC,
  long=analog-to-digital converter
}
\DeclareAcronym{AMP}{
  short=AMP,
  long=approximate message passing
}
\DeclareAcronym{ANN}{
  short=ANN,
  long=artificial neural network
}
\DeclareAcronym{AoA}{
  short=AoA,
  long=angle-of-arrival
}
\DeclareAcronym{AoD}{
  short=AoD,
  long=angle-of-departure
}
\DeclareAcronym{APS}{
  short=APS,
  long=azimuth power spectrum
}
\DeclareAcronym{AR}{
  short=AR,
  long=augmented reality
}
\DeclareAcronym{AV}{
  short=AV,
  long=autonomous vehicle
}
\DeclareAcronym{BM}{
  short=BM,
  long=beam management
}
\DeclareAcronym{BS}{
  short=BS,
  long=base station
}
\DeclareAcronym{BSM}{
  short=BSM,
  long=basic safety message
}
\DeclareAcronym{CDF}{
  short=CDF,
  long=cumulative distribution function
}
\DeclareAcronym{CP}{
  short=CP,
  long=cyclic-prefix
}
\DeclareAcronym{CSI-RS}{
  short=CSI-RS,
  long=Channel State Information Reference Signal
}
\DeclareAcronym{CSI}{
  short=CSI,
  long=Channel State Information
}
\DeclareAcronym{DFT}{
  short=DFT,
  long=discrete Fourier transform
}
\DeclareAcronym{EKF}{
  short=EKF,
  long=extended Kalman filter
}
\DeclareAcronym{DSRC}{
  short=DSRC,
  long=dedicated short-range communication
}
\DeclareAcronym{FDD}{
  short=FDD,
  long=frequency division duplex
}
\DeclareAcronym{FMCW}{
  short=FMCW,
  long=frequency modulated continuous wave
}
\DeclareAcronym{FoV}{
  short=FoV,
  long=field-of-view
}
\DeclareAcronym{GNSS}{
  short=GNSS,
  long=global navigation satellite system
}
\DeclareAcronym{IA}{
  short=IA,
  long=initial access
}
\DeclareAcronym{IMU}{
  short=IMU,
  long=inertial measurement unit
}
\DeclareAcronym{LiDAR}{
  short=LiDAR,
  long=light detection and ranging
}
\DeclareAcronym{LOS}{
  short=LOS,
  long=line-of-sight
}
\DeclareAcronym{LPF}{
  short=LPF,
  long=low pass filter
}
\DeclareAcronym{LTE}{
  short=LTE,
  long=long term evolution
}
\DeclareAcronym{MIMO}{
  short=MIMO,
  long=multiple-input multiple-output
}
\DeclareAcronym{ML}{
  short=ML,
  long=machine learning
}
\DeclareAcronym{mmWave}{
  short=mmWave,
  long=millimeter wave
}
\DeclareAcronym{MRR}{
  short=MRR,
  long=medium range radar
}
\DeclareAcronym{NLOS}{
  short=NLOS,
  long=non-line-of-sight
}
\DeclareAcronym{NB}{
  short=NB,
  long=narrow beam
}
\DeclareAcronym{NR}{
  short=NR,
  long=new radio
}
\DeclareAcronym{OFDM}{
  short=OFDM,
  long=orthogonal frequency-division multiplexing
}
\DeclareAcronym{ppm}{
  short=ppm,
  long=parts-per-million
}
\DeclareAcronym{PF}{
  short=PF,
  long=particle filter
}
\DeclareAcronym{RMS}{
  short=RMS,
  long=root-mean-square
}
\DeclareAcronym{RPE}{
  short=RPE,
  long=relative precoding efficiency
}
\DeclareAcronym{RS}{
  short=RS,
  long=reference signal
}
\DeclareAcronym{RSRP}{
  short=RSRP,
  long=reference signal received power
}
\DeclareAcronym{RSU}{
  short=RSU,
  long=roadside unit
}
\DeclareAcronym{SCS}{
  short=SCS,
  long=subcarrier spacing
}
\DeclareAcronym{SNR}{
  short=SNR,
  long=signal-to-noise ratio
}
\DeclareAcronym{SRS}{
  short=SRS,
  long=Sounding Reference Signal
}

\DeclareAcronym{SSB}{
  short=SSB,
  long=Synchronization Signal Block
}
\DeclareAcronym{THz}{
  short=THz,
  long=terahertz
}
\DeclareAcronym{UAV}{
  short=UAV,
  long=unmanned aerial vehicle
}
\DeclareAcronym{UE}{
  short=UE,
  long=user equipment
}
\DeclareAcronym{UKF}{
  short=UKF,
  long=unscented Kalman filter
}
\DeclareAcronym{ULA}{
  short=ULA,
  long=uniform linear array
}

\DeclareAcronym{UPA}{
  short=UPA,
  long=uniform planar array
}

\DeclareAcronym{V2I}{
  short=V2I,
  long=vehicle-to-infrastructure
}
\DeclareAcronym{V2V}{
  short=V2V,
  long=vehicle-to-vehicle
}
\DeclareAcronym{V2X}{
  short=V2X,
  long=vehicle-to-everything
}
\DeclareAcronym{VR}{
  short=VR,
  long=virtual reality
}
\DeclareAcronym{XR}{
  short=XR,
  long=extended reality
}
\DeclareAcronym{VRU}{
  short=VRU,
  long=vulnerable road user
}
\DeclareAcronym{WB}{
  short=WB,
  long=wide beam
}
\DeclareAcronym{BF}{
  short=BF,
  long=beamforming
}
\DeclareAcronym{CS}{
  short=CS,
  long=compressive sensing
}
\DeclareAcronym{NN}{
  short=NN,
  long=neural network
}
\DeclareAcronym{RF}{
  short=RF,
  long=radio frequency
}
\DeclareAcronym{MISO}{
  short=MISO,
  long=multiple-input single-output
}
\DeclareAcronym{SIMO}{
  short=SIMO,
  long=single-input multiple-output
}
\DeclareAcronym{MLP}{
  short=MLP,
  long=multilayer perceptron
}
\DeclareAcronym{AMCF}{
  short=AMCF,
  long=alternative minimization method with a closed-form expression
}
\DeclareAcronym{AWGN}{
  short=AWGN,
  long=additive white Gaussian noise
}
\DeclareAcronym{t-SNE}{
  short=t-SNE,
  long=t-distributed stochastic neighbor embedding
}
\DeclareAcronym{TDD}{
  short=TDD,
  long=time division duplex
}
\DeclareAcronym{mMTC}{
  short=mMTC,
  long=massive Machine Type Communications
}
\DeclareAcronym{NMSE}{
  short=NMSE,
  long=normalized mean square error
}
\DeclareAcronym{MSE}{
  short=MSE,
  long=mean square error
}
\DeclareAcronym{CNN}{
  short=CNN,
  long=convolutional neural network
}
\DeclareAcronym{Tx}{
  short=Tx,
  long=transmit
}
\DeclareAcronym{Rx}{
  short=Rx,
  long=receive
}
\DeclareAcronym{GF}{
  short=GF,
  long=grid-free
}
\DeclareAcronym{CB}{
  short=CB,
  long=codebook-based
}
\DeclareAcronym{DL}{
  short=DL,
  long=deep learning
}
\DeclareAcronym{OOB}{
  short=OOB,
  long=out-of-band
}
\DeclareAcronym{CI}{
  short=CI,
  long=context information
}
\DeclareAcronym{RL}{
  short=RL,
  long=reinforcement learning
}
\DeclareAcronym{MRT}{
  short=MRT,
  long=maximum ratio transmission
}
\DeclareAcronym{MRC}{
  short=MRC,
  long=maximum ratio combining
}
\DeclareAcronym{EGT}{
  short=EGT,
  long=equal gain transmission
}
\DeclareAcronym{EGC}{
  short=EGC,
  long=equal gain combining
}
\DeclareAcronym{DNN}{
  short=DNN,
  long=deep neural network
}
\DeclareAcronym{ReLU}{
  short=ReLU,
  long=rectified linear unit
}
\DeclareAcronym{PSD}{
  short=PSD,
  long=power spectral density
}
\DeclareAcronym{BA}{
  short=BA,
  long=beam alignment
}
\DeclareAcronym{GPU}{
  short=GPU,
  long=graphics processing unit
}
\DeclareAcronym{TCI}{
  short=TCI,
  long=transmission configuration indicator
}
\DeclareAcronym{NSA}{
  short=NSA,
  long=non-standalone
}
\DeclareAcronym{RNN}{
  short=RNN,
  long=non-recurrent neural network
}
\DeclareAcronym{LSTM}{
  short=LSTM,
  long=long short-term memory
}
\DeclareAcronym{ESI}{
  short=ESI,
  long=environmental side information
}
\DeclareAcronym{SSBA}{
  short=SSBA,
  long=site-specific beam alignment,
  first-long-format=\itshape
}
\DeclareAcronym{OTA}{
  short=OTA,
  long=over the air
}